\documentclass[11pt]{article} 

\usepackage[utf8]{inputenc} 

\usepackage{geometry} 
\usepackage{graphicx}

\usepackage{booktabs}
\usepackage{array} 
\usepackage{paralist} 
\usepackage{verbatim}
\usepackage{subfig}
\usepackage{physics}
\usepackage{amsmath,amssymb}
\usepackage{longtable}
\usepackage{stmaryrd}
\usepackage{extarrows}
\usepackage{xcolor}
\usepackage{qcircuit}
\usepackage{hyperref}

\usepackage{fancyhdr}
\usepackage{sectsty}
\allsectionsfont{\sffamily\mdseries\upshape}

\usepackage[nottoc,notlof,notlot]{tocbibind} 
\usepackage[titles,subfigure]{tocloft} 

\title{Two new non-equivalent three-qubit CHSH games}
\author{Hamza Jaffali$^{1}$, Fr\'ed\'eric Holweck$^{2,3}$}
\date{%
$^1$ \textit{ColibrITD, 91 Rue du Faubourg Saint Honor\'e, 75008 Paris, France}\\
    $^2$ \textit{Laboratoire Interdisciplinaire Carnot de Bourgogne,  UMR 6303 CNRS, University of Technology of Belfort-Montbéliard, 90010 Belfort Cedex, France}\\%
    $^3$ \textit{Mathematics and Statistics Department, Auburn University, Auburn, AL, USA}
} 

\begin{document}
\maketitle
\abstract{In this paper, we generalize to three players the well-known CHSH quantum game. To do so, we consider all possible 3 variables Boolean functions and search among them which ones correspond to a game scenario  with a quantum advantage (for a given entangled state). In particular we provide two new three players quantum games where, in one case, the best quantum strategy is obtained when the  players share a $GHZ$ state, while in the other one the players have a better advantage when they use a $W$ state as their quantum resource. To illustrate our findings we implement our game scenarios on an online quantum computer and prove experimentally the advantage of the corresponding quantum resource for each game.}
\section{Introduction}

Quantum games are games where players are allowed to use quantum resources like entanglement or quantum contextuality in order to set up winning strategies that outperform their classical counterparts. For instance, the famous prisoner dilemma which is an instance of a nonzero-sum game ceases to be a dilemma if we allow quantum strategies (see \cite{eisert1999quantum,eisert2000quantum,flitney2002introduction}).  In early 2000, quantum games such as quantum pseudo-telepathy were introduced by Brassard {\em et al.} \cite{brassard2003multi,brassard2005quantum}. Those games were exploring,  within game scenarios, some paradoxical properties of quantum physics \cite{aravind2004quantum, cabello2001all, kunkri2007winning, renner2004quantum}. One of the most famous games in the field is the CHSH game, named after Clauser, Horne, Shimony and Holt \cite{clauser1969proposed}, that opens the path to the realization of the famous experiences of Aspect \cite{aspect1982experimental} on the $EPR$ paradox and the non-locality of quantum physics. Its game-theoretic version is what we are focusing  on in this paper.

~

Let us recall the principle of this game: The game is composed of three  participants: the referee and two players: Alice (A) and Bob (B). A and B are not allowed to communicate during the game. The referee sends a binary question to each player: A receives a question $x$, B a question $y$, with $x,y \in \mathbb{B} = \{0,1\}$. Then, each player has to provide a binary answer in order to satisfy a specific equation. A will send back a bit $a$ to the referee, and B a bit $b$, without knowing what the other player decided to play. In the original version of the CHSH game, in order to win the game, the players must provide answers $(a,b)$ to the questions $(x,y)$ satisfying the following equation:

\begin{equation}
    x \cdot y = a \oplus b ~.
\end{equation}
Despite the fact that Alice and Bob can agree, before the game starts, on a common strategy, it is not too difficult to check that with classical resources and no communication during the game\footnote{We will denote those hypotheses by  $LR$ for Local Realism. When we will consider instead the hypothesis/rules of Quantum Mechanics, we will denote it by QM.}, they can only win the game with probability $p_{LR}\leq 0.75$. However, if they are allowed to use quantum resources, they can prepare an $EPR$ state, $\ket{EPR}=\dfrac{\ket{00}+\ket{11}}{\sqrt{2}}$, share its components, i.e. one qubit for Alice, one qubit for Bob and  then apply the following strategy when the game starts:
\begin{itemize}
    \item If $x=0$, A measures her qubit in the $Z$-basis and returns $0$ for a $+1$ measurement and $1$ otherwise,
    \item If $x=1$, A measures her qubit in the $X$-basis and returns $0$ for a $+1$ measurement and $1$ otherwise,
    \item If $y=0$, B measures his qubit in the $\dfrac{Z+X}{\sqrt{2}}$ basis and returns $0$ for a $+1$ measurement and $1$ otherwise,
    \item If $y=1$, B measures his qubit in the $\dfrac{Z-X}{\sqrt{2}}$ basis and returns $0$ for a $+1$ measurement and $1$ otherwise.
\end{itemize}
This quantum strategy ensures that Alice and Bob win the game with probability $p_{QM}=\cos(\frac{\pi}{8})^2\approx 0.85>0.75$.

~

Now, that those paradoxical features have been tested experimentally, it is natural to consider them as resources for the development of quantum information and quantum processing. The presence on the cloud of online quantum computers and simulators invite us to implement, test, and generalize these types of scenario both to check quantum properties by "experiences from your laptop", but also to test these early Noisy Intermediate Scale Quantum computers (NISQ) by considering how they behave when we execute such quantum games (for NISQ implementation of contextuality-based quantum games see also \cite{kelleher2023implementing}).

~

In this paper, we try to work in both directions. First, we propose an alternative method to build CHSH-like scenarios in the three-qubit case. Then, we implement and run the game on a real quantum computer and observe the violation of classical bounds. 

~

In particular, we are capable of creating two distinct scenarios that admit a quantum strategy better than any classical one, but for which the quantum resources needed for each scenario are not the same. More precisely we find first an example of a CHSH three-qubit game where the best quantum strategy is provided with a $GHZ$ entangled state, $\ket{GHZ}=\dfrac{1}{\sqrt{2}}(\ket{000}+\ket{111})$,  while in our second example, the best quantum result is obtained with a $W$-state, $\ket{W}=\dfrac{1}{\sqrt{3}}(\ket{001}+\ket{010}+\ket{100})$. In this respect, our approach differs from \cite{borsten2013freudenthal} where there were no games with better success for the $W$-state.

~

The paper is organized as follows. In Section \ref{sec:generalization} we expose our approach to generalize the CHSH game. Our idea is to consider all possible binary equations involving $2n$ variables, $n$ variables representing a set of questions $(x_1,\dots, x_n) \in \mathbb{B}^n$ and  $n$  variables $(a_1,\dots ,a_n)\in \mathbb{B}^n$ representing a set of answers provided by the $n$ players. Given an entangled state as a shared resource for the $n$ players, one can test if for a given equation one has a quantum advantage by testing all possible quantum strategies possible with that specific resource. For $n=3$, in Section \ref{sec:3qubit}, we perform an exhaustive search to find which games offer a quantum advantage when the three players share a $GHZ$ state.  We identify 2 types of equations (games) where the three players have a probability of winning with a quantum strategy which is identical to the probability of winning the CHSH game with a $EPR$ state for the same classical bound. There are other types of games where the advantage over any classical strategy is smaller.   We also do the search when the resource is a $W$ state and provide examples of games where the use of a strategy based on $W$ beats the classical strategy and the quantum strategy obtained with a $GHZ$ state. In Section \ref{sec:mermin} one discusses the relation between our approach and the family of games known as GHZ Mermin games \cite{brassard2005quantum}. In Section \ref{sec:ibm} we implement on the IBM Quantum Platform two of our new games: one game showing an advantage using a $GHZ$ state as a resource for the players, and one game based on $W$. For each game, we prove that the experimental results produced by the online quantum computer beat the classical bound and allow us to distinguish the two resources according to the outcomes of the measurement. Finally, Section \ref{sec:conclusion} is dedicated to concluding remarks and possible extensions of our work.

~

We provide the Python source code, as well as resources and results in a dedicated repository 
\href{https://github.com/ColibrITD-SAS/publications_material/tree/main/three_qubit_CHSH_game}{(link)}.

\section{Generalization of the CHSH game}\label{sec:generalization}

In this subsection, we adapt the original CHSH game to provide a general $n$-player version of the game. Our idea is, for a given $n$,  to list all possible binary equations and compare all classical strategies and their probabilities of winning the game to a quantum strategy based on a shared $n$-qubit entangled states. We also detail the method used to evaluate and determine classical and quantum strategies for several setups of the generalized CHSH game. As an illustrative example, we recover for $n=2$ several games equivalent to the CHSH game. Among other generalizations of the CHSH game for $n$ players let us mention \cite{hoban2011non, murta2016quantum} where the game is generalized to $n$ players by classes of equations, but not all of them.

    \subsection{Definition of the game}

 We denote by $A_1, A_2, \dots, A_n$ the $n$-players. Each player $i \in \llbracket 1, n \rrbracket$ receives a question $x_i \in \mathbb{B}$ from the referee and has to provide an answer $a_i \in \mathbb{B}$. To win the game, the players have to satisfy the following general equation:

\begin{equation}
    f(x_1, x_2, \dots, x_n) = g(a_1, a_2, \dots, a_n) ~.
\end{equation}

with $f,g : \mathbb{B}^n \to \mathbb{B} $, boolean functions such that each term $x_i$, respectively $a_i$, appears at least one time in the expression $f$, respectively $g$ (to assure that every player is fully involved in the game).

We denote by $h$ the strategy that the players will apply to win the game, defined as 

\begin{equation}
h \colon \biggl\{\begin{array}{@{}r@{\;}l@{}}
    \mathbb{B}^n  & \to \mathbb{B}^n \\
    (x_1, x_2, \dots, x_n) & \mapsto (a_1, a_2, \dots, a_n)
  \end{array} ~,
\end{equation}

which tells us what players will give as answers, depending on their questions. 

In other terms, the goal of the players is to determine a strategy $h$ that satisfies as much as possible the equation 

\begin{equation}
    f(x_1, x_2, \dots, x_n) = g(h(x_1, x_2, \dots, x_n)) ~.
\end{equation}

    \subsection{Finding the best classical strategy}

In the case of a classical deterministic strategy, each player will decide in advance the answer to give to the question he receives. Therefore, the strategy $h$ can be decomposed as follows:

\begin{equation}
    h(x_1, x_2, \dots, x_n) = \big(h_1(x_1), h_2(x_2), \dots, h_n(x_n) \big)
\end{equation}

with $h_i$ the strategy for the $i$-th player. For each player $i$, defining a deterministic strategy requires to define $h_i(0)$ and $h_i(1)$. Therefore, we need $2n$ bits to encode any $n$-player strategy $h$, and this implies a total of $2^{2n}$ possible classical strategies. To determine the best strategy, we generate all of them, evaluate for each the probability of gain, and keep the strategies that provide the best performance to the game.

    \subsection{Finding the best quantum strategy}

In the case of a quantum strategy, the players share a $n$-qubit quantum state $\ket{\psi_n}$. To leverage the advantage of quantum mechanics, $\ket{\psi_n}$ has to be an entangled state, otherwise, we fall back in the case of a stochastic classical strategy. Each player can manipulate his qubit by applying unitary operations. Depending on the question one player receives, he will apply one of two possible operations defined by his strategy. We denote by $U_{i,x_i}$ the unitary gate applied by the player $i$ on his qubit when he receives the question $x_i$. After all the players applied their strategy, they measure their qubit and answer the result of the measure. We summarize this process in Figure \ref{fig:quantum_strategy}.

\begin{figure}[!h]
   \begin{equation*}
    (x_1, x_2, \dots, x_n) ~ \xlongrightarrow{\text{apply strategy}} ~ U_{1,x_1} \otimes \cdots \otimes U_{n,x_n} \ket{\psi_n} ~ \xlongrightarrow{\text{measure}} ~ (a_1, a_2, \dots, a_n)
    \end{equation*}
    \caption{Overview of the application of a $n$-player quantum strategy}
    \label{fig:quantum_strategy}
\end{figure}

Any unitary operator $U_{i,x_i}$ acting on one qubit can be parametrized by three angles $\theta_{i,x_i}$, $\phi_{i,x_i}$ and $\lambda_{i,x_i}$, such that:

\begin{equation}
    U_{i,x_i}(\theta_{i,x_i}, \phi_{i,x_i},\lambda_{i,x_i}) = 
    \begin{pmatrix}
    \cos\left(\frac{\theta_{i,x_i}}{2}\right) &  -e^{j\lambda_{i,x_i}} \sin\left(\frac{\theta_{i,x_i}}{2}\right) \\
    e^{j\phi_{i,x_i}}  \sin\left(\frac{\theta_{i,x_i}}{2}\right) &        e^{j(\phi_{i,x_i} + \lambda_{i,x_i})} \cos\left(\frac{\theta_{i,x_i}}{2}\right)
    \end{pmatrix}
\end{equation}

with $j$ the complex imaginary unit.

~

To compute the performance of a quantum strategy, one has to look at the probability of measuring each basis state, after applying all the unitaries to the state $\ket{\psi_n}$. For a given question, we apply the corresponding unitaries, and we look at the analytic expression of the resulting state. Then for each basis state, we transform it into an answer by taking its binary notation, and we check if it satisfies the game's equation. We do that for each basis state, and we sum up the probabilities of winning. This gives us the probability of winning the game for a specific question. We repeat the same process for each question, and we take the mean value of probabilities of winning for all questions (we supposed that the questions are generated following a uniform distribution). This gives us the probability of winning for the overall $n$-qubit CHSH game. 

~

Therefore, finding the best quantum strategy consists of finding the angles defining the unitaries that provide the best probability of winning. For each player, 6 angles have to be found (three for each unitary), leading to $6n$ angles to optimize. We chose to reformulate the problem into a minimization problem, where the objective function $F$ is defined as : 

\begin{equation}
    F(\overrightarrow{\theta}) = 1 - P(\overrightarrow{\theta})
\end{equation}

with $ \overrightarrow{\theta} = (\theta_{1,0}, \phi_{1,0}, \lambda_{1,0}, \dots, \theta_{n,1}, \phi_{n,1}, \lambda_{n,1})$ the vector of all angles to optimize, and $P$ the function used to compute the probability of winning the game for the angles in parameter. We leverage the \verb|scipy.optimize.minimize| Python function for numerical optimizations, using \verb|BFGS| and \verb|Cobyla| methods. 

    \subsection{Example: other 2-qubit CHSH games}\label{sec:example_2qubit}

In this section, we apply the method presented in previous subsections to study a more general definition of the 2-qubit CHSH game. More precisely, we want to determine all the games (i.e. equations) allowing the players to reach a probability of gain of (at least) $|\cos(\frac{\pi}{8})|^2 \approx  0.8536$ with a quantum strategy, and for which the best classical strategy still gives 0.75 chance of winning.

~

We recall that, in this context, the two boolean functions $f$ and $g$ defining the equation to solve are functions from $\mathbb{B}^2$ to $\mathbb{B}$. Thus, we count $2^4=16$ possible functions for each of them, leading to a number of equations equal to $16 \times 16 = 256$. However, not all these equations are useful in the context of a CHSH-like game, and we need to filter them. For that, we list all possible combinations of $f$ and $g$ and only keep the ones letting all game variables $x,y$ and $a,b$ appear in the equation. In the case of 2-qubits, this is equivalent to removing the constant functions and the functions $f(x,y)=x,y,\overline{x},\overline{y}$ and $g(a,b)=a,b,\overline{a},\overline{b}$.
~

We then determine the best classical and quantum strategies of these remaining games, and only keep the ones for which quantum strategies can provide an advantage. We found a total of 16 equations providing a quantum advantage, where the user increases his probability of winning from $\frac{3}{4}$ (classical) to $|\cos(\frac{\pi}{8})|^2$ (quantum) (including the original CHSH game). Note that there is no strategy with a better probability of winning. We regroup all these equations in Table \ref{tab:function_f_2qubits}. 

\begin{table}[h!] 
\centering
 \begin{center}
  \begin{tabular}{|c|}
\hline

Equation  \\
\hline
$x \cdot y = a \oplus b$ \\
$x\cdot y = \overline{a} \oplus b$ \\
$x\cdot \overline{y} = a \oplus b$ \\
$x \cdot \overline{y} = \overline{a} \oplus b$ \\
$\overline{x} \cdot y = a \oplus b$ \\
$\overline{x} \cdot y = \overline{a} \oplus b$ \\
$\overline{x} \cdot \overline{y} = a \oplus b$ \\
$\overline{x} \cdot \overline{y} = \overline{a} \oplus b$ \\
$x + y = a \oplus b$ \\
$x + y = \overline{a} \oplus b$ \\
$x + \overline{y} = a \oplus b$ \\
$x + \overline{y} = \overline{a} \oplus b$ \\
$\overline{x} + y = a \oplus b$ \\
$\overline{x} + y = \overline{a} \oplus b$ \\
$\overline{x} + \overline{y} = a \oplus b$ \\
$\overline{x} + \overline{y} = \overline{a} \oplus b$ \\
\hline
 \end{tabular}
 \end{center}
\caption{List of all equations leading to a winning probability of $|\cos(\frac{\pi}{8})|^2$, for a classical maximum at $\frac{3}{4}$.}
\label{tab:function_f_2qubits}
\end{table}

One can reconstruct these equations by taking the four monomials $xy, x\overline{y}, \overline{x}y$ and $\overline{x}\cdot\overline{y}$ for the left-hand side, and the expression $a\oplus b$ for the right-hand side. This provides four different equations. Then one can take the conjugate of the left-hand side, to generate four other equations, take the conjugate of the right-hand side to generate four more, and take both left-hand and right-hand side conjugates. This leads us to 16 equations in total. In the case of $n=2$, our approach didn't produce any new game. Moreover all equations of Table \ref{tab:function_f_2qubits} can be deduced from the initial CHSH equation $x\cdot y=a\oplus b$ by using the logical conjugate $z\to \bar{z}$. In other words, one concludes that there is no other $2$-qubit CHSH-like game with the same quantum advantage as the original one.

\section{The 3-qubit CHSH game}\label{sec:3qubit}

In this section, we propose two new, non-equivalent, versions of the CHSH game for three players. We recall the setup of the game in the case of three qubits, and then we present our two non-equivalent games that lead to the same quantum advantage, over classical strategies, as in the 2-qubit game. 

    \subsection{Recalling the setup}

Let us call Alice (A), Bob (B), and Charlie (C) the three players. In the literature, people have mostly studied the 3-player adaptation of the CHSH game by playing three different original CHSH games (A with B, B with C, and A with C) at the same time \cite{Munoz,Delft}. We can also mention a variant called the GHZ-game, where, unlike our approach, the set of possible questions is selected (see Section \ref{sec:mermin}). Here, as explained in the general case in Section \ref{sec:generalization}, the three players play the same game, share a unique 3-qubit state, and try to satisfy the same equation.

\begin{figure}[!h]
    \centering
    \includegraphics[width=0.5\textwidth]{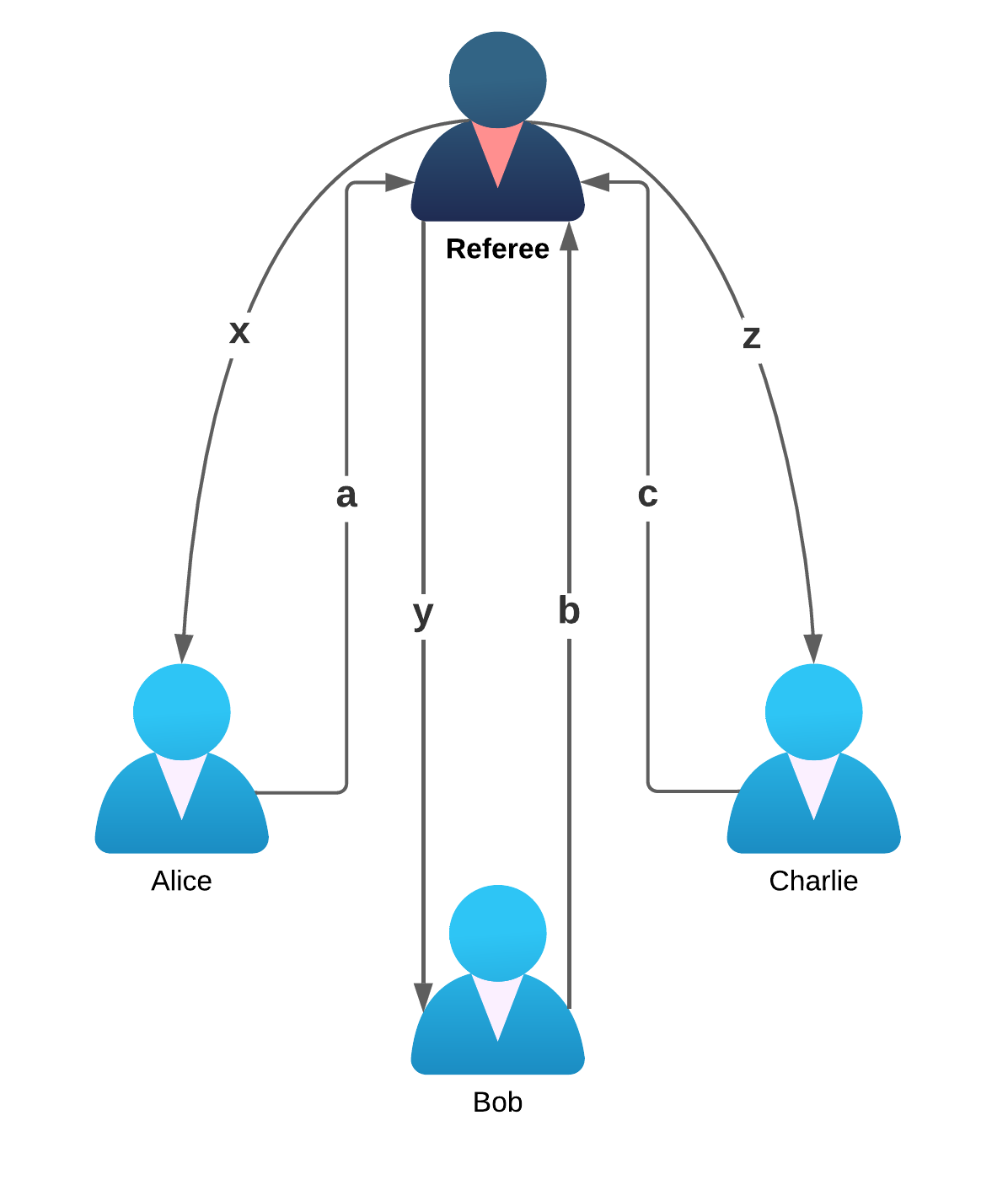}
    \caption{CHSH game setup with 3 players: The Referee sends a question $x$ to Alice, $y$ to Bob, and $z$ to Charlie. The players win the game if and only if their answers $(a,b,c)$ satisfy the binary equation $f(a,b,c)=g(x,y,z)$ that defines the game.}
    \label{fig:chsh_3}
\end{figure}

The referee sends respectively the binary questions $x$, $y$, and $z$, to A, B, and C, they apply their strategy, and they answer respectively $a$, $b$, and $c$ (see Figure \ref{fig:chsh_3}). The gain is then deduced from the satisfaction of the equation of the game, which should take the following form:

\begin{equation}\label{eq:3q-problem}
    f(x,y,z) = g(a,b,c)
\end{equation}

As recalled in Section \ref{sec:example_2qubit}, one needs to filter the useful equations among all the possible ones. We listed all the $2^8 = 256$ possible functions from $\mathbb{B}^3$ to $\mathbb{B}$ and removed 38 non-relevant ones. We ended up with $(256-38)^2= 47524$ possible games to explore in the 3-player setup. 

    \subsection{Reproducing original gains with $\ket{GHZ}$}

Each equation in the form of Equation \ref{eq:3q-problem} defines a problem to solve in a CHSH 3-player game. After studying all the possible equations, while restricting ourselves to the case where the three players share the $\ket{GHZ}$ state, and computing both the best classical and quantum gains, we particularly focus on games that provide the largest gap between classical and quantum gains. We found out that this largest gap is obtained in the three qubit game when the probability of winning with the best classical strategy is $0.75$ and $\approx 0.8535$ for the best quantum strategy. Note that this gap is the same as the one obtained in the orignial CHSH game. 

~

For all useful problems, we computed the best classical and quantum strategies, looked at the corresponding gain, and kept only the wanted ones. We counted in total 80 problems leading to the same gains as in the CHSH game. We distinguish two types of equations as follows : 

\begin{equation}\label{eq:2eq}
    \begin{split}
        \text{First type : } x  y  z + \overline{x} \cdot \overline{y} \cdot \overline{z} = a \oplus b \oplus c \\
        \text{Second type : } x y + (x \oplus y) z = a \oplus b \oplus c
    \end{split}
\end{equation}

For the first type of problem, the left-hand side is the sum (logical disjunction) of a monomial in $x$, $y$, and $z$ with the complement monomial (take the opposite of how $x$, $y$ and $z$ appear in the first monomial). There exist 8 possible monomials, and thus only 4 possible pairs. The right-hand side is the exclusive disjunction (XOR) of the three answers $a$, $b$ and $c$. The quantum and classical gains of a problem still hold if we take the negation of one of the two sides of the equations, or both at the same time. Therefore, on the left side, if we take the negation of the 4 possible sum pairs, we have 8 possible expressions. On the right side, the exclusive disjunction ($a \oplus b \oplus c$) and its negation (that can be written $\overline{a} \oplus b \oplus c$) constitute 2 expressions. Therefore, there are 16 possible problems of the first type. 

~

For the second type of problem, we count 64 equations in total. Starting from the 16 equations listed in Table \ref{tab:second_type_GHZ_wins}, one can take the negation of the left-hand side of each equation to generate 16 other equations. Then, one can also take the negation of the right-hand side of each of these 32 equations, retrieving all 64 equations.

\begin{table}[h!]
    \centering
    \begin{tabular}{|c|c|}
        \hline 
        $xy + (x \oplus y)z = a \oplus b \oplus c$ & $xy + (x \oplus y)\bar{z} = a \oplus b \oplus c$ \\
        \hline 
        $x\bar{y} + (x \oplus z)y = a \oplus b \oplus c$ & $x\bar{y} + (\bar{x} \oplus z)y = a \oplus b \oplus c$ \\
        \hline
        $\bar{x}y + (y \oplus z)x = a \oplus b \oplus c$ & $\bar{x}y + (\bar{y} \oplus z)x = a \oplus b \oplus c$ \\
        \hline
        $xy + (x \oplus z)\bar{y} = a \oplus b \oplus c$ & $x\bar{y} + (\bar{x} \oplus y)z = a \oplus b \oplus c$ \\
        \hline
        $\bar{x}z + (y \oplus z)x = a \oplus b \oplus c$ & $\bar{x}z + (\bar{y} \oplus z)x = a \oplus b \oplus c$ \\
        \hline
        $xy + (y \oplus z)\bar{x} = a \oplus b \oplus c$ & $xz + (y \oplus z)\bar{x} = a \oplus b \oplus c$ \\
        \hline
        $x\bar{z} + (y \oplus z)\bar{x} = a \oplus b \oplus c$ & $x\bar{y} + (y \oplus z)\bar{x} = a \oplus b \oplus c$ \\
        \hline
        $\bar{x}y + (\bar{x} \oplus y)z = a \oplus b \oplus c$ & $\bar{x}y + (x \oplus z)\bar{y} = a \oplus b \oplus c$ \\
        \hline
    \end{tabular}
    \caption{List of 16 equations that can be used to generate all 64 equations of the second type.}
    \label{tab:second_type_GHZ_wins}
\end{table}

We present in Table \ref{tab:examples_GHZ_strategies} an example of optimal quantum strategies for one example of each type of equation. We first observe that, in both of these examples, the three players can apply the same quantum strategies to reach the maximum gain. This can be explained by the symmetry present in both equations, but also in the $\ket{GHZ}$ state. We also point out remarkable values for strategies' angles: $\forall (i,j) \in \{0,1,2\} \times \mathbb{B}, ~ \theta_{i,j}= \frac{\pi}{2}, ~\phi_{i,j}=0$ and $\forall j \in \mathbb{B}, ~ | \lambda_{i,0} - \lambda_{i,1}| = \frac{\pi}{2}$.   

\begin{table}[h!]
    \centerline{
    \begin{tabular}{|c|c|c|c|c|c|c|}
        \hline 
        Equation example & $U_{1,0}$ & $U_{1,1}$ & $U_{2,0}$ & $U_{2,1}$ & $U_{3,0}$ & $U_{3,1}$ \\
        \hline 
        \hline 
        $x y z + \overline{x} \cdot \overline{y} \cdot \overline{z} = a \oplus b \oplus c$ & $(\frac{\pi}{2},0,\frac{\pi}{12})$ & $(\frac{\pi}{2},0,\frac{7\pi}{12})$ & $(\frac{\pi}{2},0,\frac{\pi}{12})$ & $(\frac{\pi}{2},0,\frac{7\pi}{12})$& $(\frac{\pi}{2},0,\frac{\pi}{12})$ & $(\frac{\pi}{2},0,\frac{7\pi}{12})$ \\
        \hline

        $x y + (x \oplus y) z = a \oplus b \oplus c$ & $(\frac{\pi}{2},0,\frac{-\pi}{4})$ & $(\frac{\pi}{2},0,\frac{-3\pi}{4})$ & $(\frac{\pi}{2},0,\frac{-\pi}{4})$& $(\frac{\pi}{2},0,\frac{-3\pi}{4})$ & $(\frac{\pi}{2},0,\frac{-\pi}{4})$ & $(\frac{\pi}{2},0,\frac{-3\pi}{4})$ \\
        \hline
    \end{tabular}}
    \caption{Quantum strategies for two different scenarios (Equations (\ref{eq:2eq})) of three players game where the maximum advantage is achieved for a $\ket{GHZ}$ state. With the quantum strategy, the players win with probability $\approx 0.85$ compared to $0.75$ for all classical strategies.}
    \label{tab:examples_GHZ_strategies}
\end{table}

    \subsection{Performance with $\ket{W}$ state}

In the 3-qubit case, we know that there exist two non-equivalent classes of genuinely entangled states \cite{Gour2000}, represented by the $\ket{GHZ}$ and $\ket{W}$ states. A natural question is then to study the performance of the players when they share a $\ket{W}$ state instead of a $\ket{GHZ}$ one, and to compare the corresponding gains. 

~

We thus explored all possible equations where the classical gain is fixed to 0.75, and for which sharing the state $\ket{W}$ provides an advantage. We then compute the performance with the $\ket{GHZ}$ for the same game. For several equations of the same type, we retrieve the same results with both quantum states. We report in Table \ref{tab:examples_W_results} only a single example for each game, with the corresponding gains for both $\ket{W}$ and $\ket{GHZ}$.


\begin{table}[h!]
    \centering
    \begin{tabular}{|c|c|c|c|}
        \hline 
        Equation example & gain$_\text{classical}$ & gain$_{\ket{W}}$  & gain$_{\ket{GHZ}}$ \\
        \hline 
        \hline 
        $yz + x\bar{z} = \bar{a}bc + a\bar{b}c + ab\bar{c}$ & 0.75 & 0.75442 & 0.69887 \\
        \hline 
        $(x \oplus y)z + xy = \bar{a} \oplus b \oplus c $ & 0.75 & 0.77216 & 0.85355\\
        \hline
        $x(y\oplus z) = (a \oplus b)c$ & 0.75 & 0.77523 & 0.80177 \\
        \hline
        $\bar{x}yz + x\bar{y}\bar{z} = \bar{a}\bar{b}\bar{c} + \bar{a}bc + a\bar{b}c + ab\bar{c} + abc$ & 0.75 & 0.77674 & 0.70266 \\
        \hline
        $(x \oplus y)z + xy = (a \oplus b)\bar{c}+ ac$ & 0.75 & 0.78726 & 0.75 \\
        \hline
        $\bar{x}yz + x\bar{y}\bar{z} = a ( b \oplus c)$ & 0.75 & 0.79219 & 0.80177 \\
        \hline
        $\bar{x}yz + x\bar{y}\bar{z} = \bar{a}bc + a\bar{b}c + ab\bar{c}$ & 0.75 & 0.79665 & 0.82766 \\
        \hline 
        $ (x \oplus y)z + x\bar{z} = a \oplus b \oplus c$ & 0.75 & 0.80046 & 0.85355 \\
        \hline
    \end{tabular}
    \caption{Example of equations for which the best quantum strategy with the $\ket{W}$ state outperform the best classical strategy (0.75 of gain). The best quantum strategy with $\ket{GHZ}$ is also presented to compare.}
    \label{tab:examples_W_results}
\end{table}

~

In most cases listed in the table, sharing a $\ket{GHZ}$ state allows the players to reach a higher gain than with the $\ket{W}$ state, while still outplaying all classical strategies with both states. However, for some games, at least three of them, $\ket{W}$ shows better performance than $\ket{GHZ}$ state, with $\ket{GHZ}$ providing a gain that is lower or equal to the classical gain. This allows one to detect a $\ket{W}$ state by detecting a violation of the classical upper bound and thus provides a way to characterize and differentiate between 3-qubit genuine types of entanglement.

\begin{table}[h!]
    \centering
    \begin{tabular}{|c|c|c|}
        \hline 
        Equation example & $U_{1,0}$ & $U_{1,1}$ \\
        \cline{1-2} 
        \hline
         & $(2.3177324, 0, \frac{-5\pi-2}{6})$ & $(2.3177324, 0, \frac{\pi-2}{6})$\\
        \cline{2-3}
        \cline{2-3}
        &  $U_{2,0}$ & $U_{2,1}$ \\
        \cline{2-3} 
        $xyz + \overline{x}\cdot\overline{y}\cdot\overline{z} = \overline{a}\overline{b}c + \overline{a}bc + a\overline{b}\overline{c} + abc$ &  $(0.8238602, 0, \frac{\pi-2}{6})$ & $(-0.8238602, 0, \frac{\pi-2}{6})$ \\
        \cline{2-3} 
        \cline{2-3} 
        & $U_{3,0}$ & $U_{3,1}$ \\
        \cline{2-3} 
           & $(0.79655904, 0, \frac{\pi-2}{6})$ & $(-0.79655904, 0, \frac{\pi-2}{6})$ \\
        \cline{2-3}
        \hline
    \end{tabular}
    \caption{Quantum strategies for three players game where the maximum advantage is achieved with a $\ket{W}$ state. With the quantum
 strategy, the players win with a probability $\approx$ 0.78726 compared to 0.75 with $\ket{GHZ}$ state and all classical strategies.}
    \label{tab:examples_W_strategies}
\end{table}

\section{Connection with Mermin's  inequalities}\label{sec:mermin}

The CHSH game was first introduced as a possible experimental protocol to test Bell's inequalities. In this section, we would like to recall the connection between Bell's operators and the CHSH game and show how our approach differs from another possible generalization known as the GHZ-game \cite{mermin1990quantum,brassard2003multi}.

~

Recall the following  expression of Bell's operator:

\begin{equation}\label{bellop}
\mathcal{B}=Z\otimes (\frac{Z+X}{\sqrt{2}})+X\otimes (\frac{Z+X}{\sqrt{2}})+ Z\otimes (\frac{Z-X}{\sqrt{2}})-X\otimes (\frac{Z-X}{\sqrt{2}}) ~.
\end{equation}
From the rules of Quantum mechanics, a straightforward calculation shows that,
\begin{equation}
    \langle \mathcal{B}\rangle_{\ket{EPR}}=2\sqrt{2}.
\end{equation}
However, as proven by Bell \cite{bell1964einstein}, any LR theory attempting to assign pre-definite values to the observables $Z, X, \dfrac{Z+X}{\sqrt{2}}, \dfrac{Z-X}{\sqrt{2}}$ would lead to another value bounded by $2$.  More precisely Bell's inequalities can be stated as:
\begin{equation}
\langle B\rangle \leq b \text{ where } \left\{\begin{array}{cc}
b^{QM}=2\sqrt{2}\\
b^{LR}=2
\end{array}.\right.
\end{equation}
One recognizes in Eq. (\ref{bellop}) the four observables that define the four experimental settings in the CHSH protocol. To emphasize the connection between Bell's operator and the CHSH game, let us recall that for instance when both players Alice and Bob get the question $x=y=0$ they perform the experiments corresponding to the monomial $Z\otimes \dfrac{Z+X}{\sqrt{2}}$. In this case, they win the game if they both send the same answer to Charlie, i.e. the result of the two measurements is $+1$  (either they measure $(1,1)$ and send $(a,b)=(0,0)$ or they measure $(-1,-1)$ and send $(a,b)=(1,1)$). If one denotes by $p$ the probability of winning the CHSH game in this case, i.e. of measuring $+1$ for this specific monomial one has 
\begin{equation}\label{eq:probachsh}
1\times p -1\times (1-p)=\langle Z\otimes \dfrac{Z+X}{\sqrt{2}}\rangle.
\end{equation}
With the $EPR$ states it leads to 
\begin{equation}\label{eq:proba}
2p-1=\langle Z\otimes \dfrac{Z+X}{\sqrt{2}}\rangle_{\ket{EPR}}=\frac{1}{\sqrt{2}}.
\end{equation}

In other words $p=\dfrac{1}{2}+\dfrac{1}{2\sqrt{2}}$, i.e. $p=\cos^2(\frac{\pi}{8})$ as claimed by the CHSH game (similar calculation leads to the same conclusion in the three other cases).

~

Bell's inequalities can be generalized to $n\geq 3$ by the so-called Mermin's inequalities \cite{mermin1990extreme,belinskiui1993interference}.  Mermin's inequalities are based on Mermin's operators that are defined inductively for any $n$ and contain Bell's operator as a particular case. According to the inductive process defined by Mermin, for $n=3$, a Mermin operator has the following form:
\begin{equation}
M_3=\mathcal{O}_{1,0}\otimes \mathcal{O}_{2,0}\otimes \mathcal{O}_{3,1}+\mathcal{O}_{1,0}\otimes \mathcal{O}_{2,1}\otimes \mathcal{O}_{3,0}+\mathcal{O}_{1,1}\otimes \mathcal{O}_{2,0}\otimes \mathcal{O}_{3,0}-\mathcal{O}_{1,1}\otimes \mathcal{O}_{2,1}\otimes \mathcal{O}_{3,1},
\end{equation}
where $\mathcal{O}_{i,0}$ and $\mathcal{O}_{i,1}$ are two distinguished one-qubit observables. 

~

Like Bell's operator, there are inequalities associated to Mermin's operators depending on the description of reality that is chosen. For $n=3$ one gets:
\begin{equation}
\langle M_3\rangle \leq b \text{ where } \left\{\begin{array}{cc}
b^{QM}=4\\
b^{LR}=2
\end{array}.\right.
\end{equation}
However, unlike Bell's operator, the three-qubit observables involved in $M_3$ are all mutually commuting. Therefore there exists a common eigenstate. More precisely if  $\mathcal{O}_{k,0}=X$ and $\mathcal{O}_{k,1}=Y$, for $k=1,2,3$, then the $GHZ$-like state 
\[\ket{GHZ_j}=\dfrac{1}{\sqrt{2}}(\ket{000}+j\ket{111})\]
is an eigenvector of eigenvalue $+1$ for $X\otimes X\otimes Y, X\otimes Y\otimes X, Y\otimes X\otimes X$ but $-1$ for the observable $Y\otimes Y\otimes Y$ leading to (the maximum) value:
\[\langle M_3\rangle_{\ket{GHZ_j}}=4.\]
Based on Mermin's polynomial $M_3$ one can derive a three-player game known as the GHZ game \cite{brassard2005quantum,mermin1990quantum}. In this game, the referee sends his questions $(x,y,z)\in \{0,1\}^3$ to Alice, Bob, and Charlie but one assumes that the number of $1$ in each triplet sent is odd (i.e. one only considers the set of questions $S=\{(0,0,1), (0,1,0), (1,0,0), (1,1,1)\}$. To win the game the answer $(a,b,c)\in \{0,1\}^3$ of Alice, Bob and Charlie should satisfy 
\[a\oplus b\oplus c=xyz.\]
Now the fact that $\ket{GHZ_j}$ is an eigenvector of the observables involved in $M_3$ implies that a quantum strategy consisting of measuring $\ket{GHZ_j}$ in the $X$-basis for question $0$ or in the $Y$-basis if the question is $1$ will be successful with probability one for the set of question $S$.  Indeed compared to Eq (\ref{eq:probachsh}) we have this time
\begin{equation}
2p-1=\langle X\otimes X\otimes Y\rangle_{\ket{GHZ_j}}=1.
\end{equation}
Let us remark here that $\ket{GHZ_j}$ is not an eignevector for the $M_3'$ operator i.e. the one obtained from $M_3$ by switching the role of $\mathcal{O}_{i,0}$
and $\mathcal{O}_{i,1}$, 

\begin{equation}
    M_3'=X\otimes Y\otimes Y+Y\otimes X\otimes Y+Y\otimes Y\otimes X-X\otimes X\otimes X ~,
\end{equation}

and thus a quantum strategy defined by $M_3'$ for the quantum resource $\ket{GHZ_j}$ would not have any advantage compared to a classical strategy.

~

This detour by the GHZ-Mermin games is useful to emphasize what is different in our approach. First because one considers all possible questions in our setting, the corresponding operator would be made of $8$ monomials, or in other words, one considers the operator $M=M_3\pm M_3'$. Those $8$ monomials operators do correspond to two sets of 4 mutually commuting ones. More precisely one has two sets:
\[\mathcal{S}_1=\{\mathcal{O}_{1,0}\otimes \mathcal{O}_{2,0}\otimes \mathcal{O}_{3,1},\mathcal{O}_{1,0}\otimes \mathcal{O}_{2,1}\otimes \mathcal{O}_{3,0},\mathcal{O}_{1,1}\otimes \mathcal{O}_{2,0}\otimes \mathcal{O}_{3,0},\mathcal{O}_{1,1}\otimes \mathcal{O}_{2,1}\otimes \mathcal{O}_{3,1} \}\] \[\text{ and } \mathcal{S}_2=\{\mathcal{O}_{1,1}\otimes \mathcal{O}_{2,1}\otimes \mathcal{O}_{3,0},\mathcal{O}_{1,1}\otimes \mathcal{O}_{2,0}\otimes \mathcal{O}_{3,1},\mathcal{O}_{1,0}\otimes \mathcal{O}_{2,1}\otimes \mathcal{O}_{3,1},\mathcal{O}_{1,0}\otimes \mathcal{O}_{2,0}\otimes \mathcal{O}_{3,0} \}.\]
And our quantum resource, $\ket{GHZ}$ or $\ket{W}$ are neither an eigenvector of the first or the second set but roughly speaking the best candidate to maximize the expectation of a combination of those two sets.

~

Let's be more specific: if one considers the two games provided by Table \ref{tab:examples_GHZ_strategies} that we discovered by our approach, then one can define for both games the following polynomial operator

\begin{equation}
T_1=M_3-M_3',
\end{equation}
with  $\mathcal{O}_{i,0}$ and $\mathcal{O}_{i,1}$ as in the Table \ref{tab:examples_GHZ_strategies}.
Then a quick calculation shows that 
\begin{equation}
\langle T_1\rangle_{\ket{GHZ}}=4\sqrt{2},
\end{equation}
which by construction should be an upper bound. Indeed the fact that $\ket{GHZ}$ is the resource that provides the maximum winning probabilities for the equation that defines the game implies that it also achieves the maximum of $\langle T_1\rangle$. Note that the classical bound for this operator is $0$.


~

Now if one looks at the $W$-game provided in Table \ref{tab:examples_W_strategies}, then one can show that the operator associated with our strategy is:
\begin{equation}
    T_2=M_3+M_3'.
\end{equation}
A straight calculation shows that 
$\langle T_2\rangle_{\ket{W}}\approx 3.7922$.

~

It is interesting to point out here that our approach based on an exhaustive search of binary games with a quantum advantage has led us to the expression of operators $T_1$ and $T_2$, defining Bell-like type inequalities but for different resources. Let us mention here that, from a different perspective, other new types of Bell's inequalities with peculiar properties are derived in \cite{hoban2011non} from CHSH-like games.

\section{Playing the three-qubit CHSH games with a quantum computer}\label{sec:ibm}

In this section, we propose to play two of our different 3-qubit CHSH games on real quantum devices. The first motivation is to observe the violation of classical bounds by playing our adaptation of the CHSH games. The second one is to assess the performance of the quantum protocol in a noisy setup and on a real quantum device. We selected two different games, one where the $\ket{GHZ}$ state provides an advantage over the classical gain, and a second where the $\ket{W}$ beats the classical limit and reaches a higher score than the $\ket{GHZ}$ one. In both settings, the classical bound is beaten by the quantum strategies in agreement with the theory. In the second setting, our game allows us to distinguish between the $\ket{GHZ}$ and the $\ket{W}$ state.

    \subsection{First experiment}\label{sec:game1_played}

The first game we play is defined by Equation (\ref{eq:first_experiment}), and corresponds to the second type of equations presented in Equation (\ref{eq:2eq}). Recall that, in that case, one can outperform the best classical strategy (0.75 of gain) with a quantum strategy given in Table \ref{tab:examples_GHZ_strategies} involving the $\ket{GHZ}$ state (0.8535 of gain).

\begin{equation}\label{eq:first_experiment}
    xy + (x \oplus y)z = a \oplus b \oplus c
\end{equation}

\begin{figure}[!h]
    \centerline{
    \Qcircuit @C=1em @R=.7em {
    & \gate{H} & \ctrl{1} & \qw & \gate{U_{1,x}} & \meter & \qw\\
    & \qw & \targ  & \ctrl{1} & \gate{U_{2,y}} & \meter & \qw\\
    & \qw & \qw  & \targ & \gate{U_{3,z}} & \meter & \qw \\
    }}
    \caption{Quantum circuit used for applying the quantum strategy. The optimal angles of the local unitary transformations depend on the question sent to the players and are reported in Table \ref{tab:examples_GHZ_strategies} (second line).}
    \label{fig:circuit_game_ghz}
\end{figure}
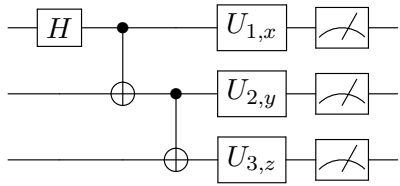

To play the game, we first generate the $\ket{GHZ}$ state by applying a Hadamard gate on the first qubit and a cascade of CNOT gates. Each player then applies a unitary transformation on his qubit, depending on the received question, with angles determined in previous sections. Each qubit is then measured in the computational basis and the result is returned to the referee. The executed circuit is represented in Figure \ref{fig:circuit_game_ghz}

\begin{figure}[!h]
    \centering
    \includegraphics[width=0.9\textwidth]{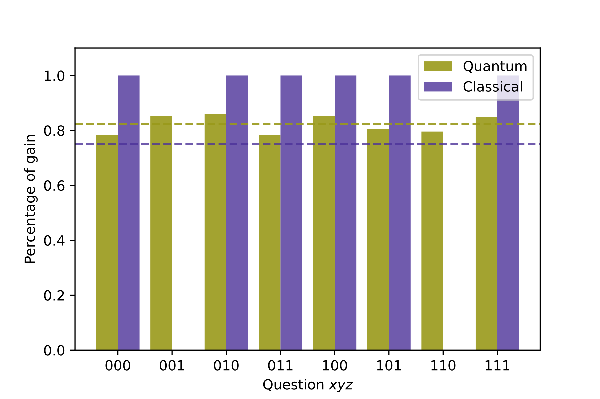}
    \caption{Histogram of the percentage of victory as a function of the question $x,y,z$ sent to the three players. In green is the percentage of victory when using the quantum strategy of the second line of  Table \ref{tab:examples_GHZ_strategies}. In purple a classical strategy that achieves the maximum bound of $75\%$ of victory. The dashed line represents the average victory. The experiment was performed on IBM\_Lagos with 10,000 shots on February 21, 2023.}
    \label{fig:game_1}
\end{figure}

We execute the game on the 5-qubit IBM\_LAGOS device. We report in Figure \ref{fig:game_1} the percentage of gain for each question, in the form of a histogram, for both best classical and quantum strategies. We first observe that the classical strategy is either completely winning or losing depending on the question asked. We also remark that every percentage of quantum gain for each question is higher than the classical limit of 0.75. Finally, we clearly observe that the quantum strategy with shared $\ket{GHZ}$ state provides an advantage over the best classical strategy.

    \subsection{Second experiment}\label{sec:game2_played}
For the second experiment, one considers the game where the players Alice, Bob, and Charlie win whenever their answers $(a,b,c)\in \mathbb{B}_2$ satisfy the following Equation (\ref{eq:2ndexp}) for a choice of question $(x,y,z)\in \mathbb{B}_2$.

\begin{equation}\label{eq:2ndexp}
    xyz + \overline{x}\cdot\overline{y}\cdot\overline{z} = \overline{a}\overline{b}c + \overline{a}bc + a\overline{b}\overline{c} + abc.
\end{equation}

This equation is nothing but the equation of Table \ref{tab:examples_W_strategies} where there is a strategy that achieves the optimal advantage using a $\ket{W}$ state.

~

\begin{figure}[!h]
    \centerline{
    \Qcircuit @C=1em @R=.7em {
    & \gate{R_y(\phi)} & \ctrl{1} & \qw & \ctrl{1} & \gate{X} & \gate{U_{1,x}} & \meter & \qw\\
    & \qw & \gate{H}  & \ctrl{1} & \targ & \qw & \gate{U_{2,y}} & \meter & \qw\\
    & \qw & \qw  & \targ &  \qw & \qw & \gate{U_{3,z}} & \meter & \qw \\
    }}
    \caption{with $\phi = 2\arccos{\frac{1}{\sqrt{3}}}$ ,}
    \label{fig:circuit_game_w}
\end{figure}
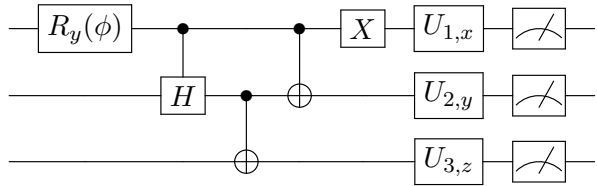

Like for the first experiment, one needs to prepare our entangled quantum resource, i.e. the $\ket{W}$ state and then for each player performs the unitary corresponding to the measurement the player has to perform for each question $x,y$ or $z$ (see Figure \ref{fig:circuit_game_w}).

~

\begin{figure}[!h]
    \centering
    \includegraphics[width=0.9\textwidth]{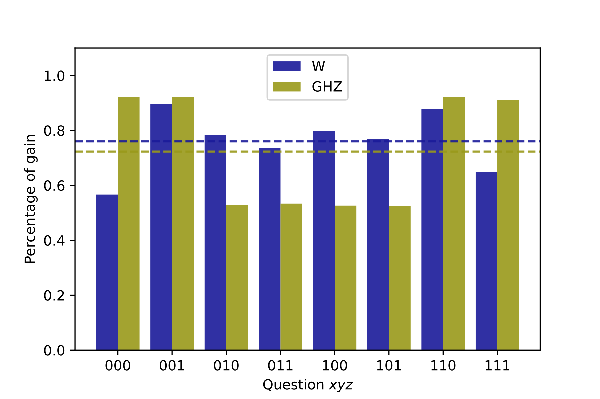}
    \caption{Histogram of the percentage of victory as a function of the question $x,y,z$ sent to the three players. In green is the percentage of victory when using the $\ket{GHZ}$ state as a quantum resource. In blue the quantum strategy involving the $\ket{W}$ state surpasses the maximum bound of $75\%$ of victory. The dashed line represents the average victory. The experiment was performed on IBM\_Lagos with 20,000 shots on February 21, 2023.}\label{fig:game_2}
\end{figure}

We ran this game on the $5$-qubit IBM\_LAGOS device. Figure \ref{fig:game_2} presents our result in the form of a histogram comparing the percentage of victory when playing the game with a $\ket{W}$ state and with a $\ket{GHZ}$ state. For one half of the question, the best strategy with the $\ket{GHZ}$ beats the best strategy with $\ket{W}$. But on average, the percentage of win is higher with $\ket{W}$ and beats the classical bound of $75\%$. In this experiment, the use of the online IBM quantum computer shows again that the classical bound can be violated with a quantum strategy but it also allows us to distinguish $\ket{W}$ from $\ket{GHZ}$. Indeed one knows that for this game the maximum probability of winning with a $\ket{GHZ}$ strategy is the same as with the best classical strategy. The fact that one wins with a percentage of 76\% shows that the state used as a resource can not be $\ket{GHZ}$. Besides, with the presence of noise, one only wins with a percentage of 72.3\% gives us a worse gain than the one we can obtain with the best classical strategy.

\section{Conclusion}\label{sec:conclusion}

In this article, we investigated the idea of generalizing the CHSH game from the perspective of the different types of binary equations that can be advantageously solved by players using an entangled state. Our approach shows that for two-qubit games there are no other equations than the original equation $a\oplus b=x.y$ for which sharing an EPR state provides a quantum advantage. However, once we consider three-player games, one can find non-equivalent games, i.e. non-equivalent binary equations, for which there is a quantum advantage in using genuine three-qubit entangled states. To illustrate this, one selected two non-equivalent games using the three-qubit entangled state $\ket{GHZ}$ as the optimal resource and one game using the state $\ket{W}$ as the optimal resource. These last three games were experimentally implemented using an online quantum computer. We showed in particular that those experiments exhibit the quantum advantage of the games and allow us to distinguish the $\ket{GHZ}$ and $\ket{W}$ states from each other. In this respect, it confirms the potential of the application of quantum games to self-testing and device-independent quantum computing \cite{murta2016quantum, pironio2016focus, vsupic2020self}.

~

The explicit enumeration of binary equations, and their test as a support of a quantum strategy based on a given quantum resource, could be used to explore also four-qubit games and/or games based on two or three-qutrits. In the four-qubit games, this could shed some new light on the question of the four-qubit classification \cite{verstraete2002four,holweck2014entanglement,holweck2017entanglement,jaffali2019quantum}. Those games are all related to some Bell-like inequalities that can be expressed in terms of operators (as discussed in Section \ref{sec:mermin}). But starting from a naive enumeration of all binary equations could bring interesting perspectives in terms of applications of quantum games. Indeed most of the future applications of quantum information are unknown today and we believe that the elementary formulation of the quantum games in terms of binary equations to solve could more easily be translated to other fields for potential applications.

\section*{Acknowledgment}
This work is supported by the Graduate school EIPHI (contract ANR-17-EURE-
0002) through the project TACTICQ. We acknowledge the use of the IBM Quantum 
Experience and the IBMQ-research program. The views expressed are those of the authors and do not reflect the official policy or position of IBM or the IBM Quantum Experience team. One would like to thank the developers of the open-source framework Qiskit.
\bibliography{main}

\begin{thebibliography}{10}

\bibitem{aravind2004quantum}
Padmanabhan~K Aravind.
\newblock Quantum mysteries revisited again.
\newblock {\em American Journal of Physics}, 72(10):1303--1307, 2004.

\bibitem{aspect1982experimental}
Alain Aspect, Jean Dalibard, and G{\'e}rard Roger.
\newblock Experimental test of bell's inequalities using time-varying
  analyzers.
\newblock {\em Physical review letters}, 49(25):1804, 1982.

\bibitem{belinskiui1993interference}
Belinski{\u\i}.
\newblock Interference of light and bell's theorem.

\bibitem{bell1964einstein}
John~S Bell.
\newblock On the einstein podolsky rosen paradox.
\newblock {\em Physics Physique Fizika}, 1(3):195, 1964.

\bibitem{borsten2013freudenthal}
L~Borsten.
\newblock Freudenthal ranks: Ghz versus w.
\newblock {\em Journal of Physics. A, Mathematical and Theoretical (Online)},
  46, 2013.

\bibitem{brassard2003multi}
Gilles Brassard, Anne Broadbent, and Alain Tapp.
\newblock Multi-party pseudo-telepathy.
\newblock In {\em Algorithms and Data Structures: 8th International Workshop,
  WADS 2003, Ottawa, Ontario, Canada, July 30-August 1, 2003. Proceedings 8},
  pages 1--11. Springer, 2003.

\bibitem{brassard2005quantum}
Gilles Brassard, Anne Broadbent, and Alain Tapp.
\newblock Quantum pseudo-telepathy.
\newblock {\em Foundations of Physics}, 35(11):1877--1907, 2005.

\bibitem{cabello2001all}
Ad{\'a}n Cabello.
\newblock “all versus nothing” inseparability for two observers.
\newblock {\em Physical Review Letters}, 87(1):010403, 2001.

\bibitem{clauser1969proposed}
John~F Clauser, Michael~A Horne, Abner Shimony, and Richard~A Holt.
\newblock Proposed experiment to test local hidden-variable theories.
\newblock {\em Physical review letters}, 23(15):880, 1969.

\bibitem{Gour2000}
W.~D\"ur, G.~Vidal, and J.~I. Cirac.
\newblock Three qubits can be entangled in two inequivalent ways.
\newblock {\em Phys. Rev. A}, 62:062314, Nov 2000.

\bibitem{eisert2000quantum}
Jens Eisert and Martin Wilkens.
\newblock Quantum games.
\newblock {\em Journal of Modern Optics}, 47(14-15):2543--2556, 2000.

\bibitem{eisert1999quantum}
Jens Eisert, Martin Wilkens, and Maciej Lewenstein.
\newblock Quantum games and quantum strategies.
\newblock {\em Physical Review Letters}, 83(15):3077, 1999.

\bibitem{flitney2002introduction}
Adrian~P Flitney and Derek Abbott.
\newblock An introduction to quantum game theory.
\newblock {\em Fluctuation and Noise Letters}, 2(04):R175--R187, 2002.

\bibitem{hoban2011non}
Matty~J Hoban, Earl~T Campbell, Klearchos Loukopoulos, and Dan~E Browne.
\newblock Non-adaptive measurement-based quantum computation and multi-party
  bell inequalities.
\newblock {\em New Journal of Physics}, 13(2):023014, 2011.

\bibitem{holweck2014entanglement}
Fr{\'e}d{\'e}ric Holweck, Jean-Gabriel Luque, and Jean-Yves Thibon.
\newblock Entanglement of four qubit systems: A geometric atlas with polynomial
  compass i (the finite world).
\newblock {\em Journal of Mathematical Physics}, 55(1), 2014.

\bibitem{holweck2017entanglement}
Fr{\'e}d{\'e}ric Holweck, Jean-Gabriel Luque, and Jean-Yves Thibon.
\newblock Entanglement of four-qubit systems: a geometric atlas with polynomial
  compass ii (the tame world).
\newblock {\em Journal of Mathematical Physics}, 58(2), 2017.

\bibitem{jaffali2019quantum}
Hamza Jaffali and Fr{\'e}d{\'e}ric Holweck.
\newblock Quantum entanglement involved in grover’s and shor’s algorithms:
  the four-qubit case.
\newblock {\em Quantum Information Processing}, 18:1--41, 2019.

\bibitem{kelleher2023implementing}
Colm Kelleher, Mohammad Roomy, and Fr{\'e}d{\'e}ric Holweck.
\newblock Implementing 2-qubit pseudo-telepathy games on noisy intermediate
  scale quantum computers.
\newblock {\em arXiv preprint arXiv:2310.07441}, 2023.

\bibitem{kunkri2007winning}
Samir Kunkri, Guruprasad Kar, Sibasish Ghosh, and Anirban Roy.
\newblock Winning strategies for pseudo-telepathy games using single non-local
  box.
\newblock {\em Quantum Information \& Computation}, 7(4):319--328, 2007.

\bibitem{mermin1990extreme}
N~David Mermin.
\newblock Extreme quantum entanglement in a superposition of macroscopically
  distinct states.
\newblock {\em Physical Review Letters}, 65(15):1838, 1990.

\bibitem{mermin1990quantum}
N~David Mermin.
\newblock Quantum mysteries revisited.
\newblock {\em American Journal of Physics}, 58(8):731--734, 1990.

\bibitem{murta2016quantum}
Gl{\'a}ucia Murta, Ravishankar Ramanathan, Nat{\'a}lia M{\'o}ller, and
  Marcelo~Terra Cunha.
\newblock Quantum bounds on multiplayer linear games and device-independent
  witness of genuine tripartite entanglement.
\newblock {\em Physical Review A}, 93(2):022305, 2016.

\bibitem{pironio2016focus}
Stefano Pironio, Valerio Scarani, and Thomas Vidick.
\newblock Focus on device independent quantum information.
\newblock {\em New J. Phys}, 18(10):100202, 2016.

\bibitem{renner2004quantum}
Renato Renner and Stefan Wolf.
\newblock Quantum pseudo-telepathy and the kochen-specker theorem.
\newblock In {\em International Symposium onInformation Theory, 2004. ISIT
  2004. Proceedings.}, pages 322--322. IEEE, 2004.

\bibitem{vsupic2020self}
Ivan {\v{S}}upi{\'c} and Joseph Bowles.
\newblock Self-testing of quantum systems: a review.
\newblock {\em Quantum}, 4:337, 2020.

\bibitem{Delft}
{TUDelft - Vidick}.
\newblock Quantum cryptography : 2.6.2 playing chsh with three players.
\newblock
  \url{https://ocw.tudelft.nl/course-lectures/2-6-2-playing-chsh-three-players/}.
\newblock Course subject(s) - 02. The power of entanglement.

\bibitem{verstraete2002four}
Frank Verstraete, Jeroen Dehaene, Bart De~Moor, and Henri Verschelde.
\newblock Four qubits can be entangled in nine different ways.
\newblock {\em Physical Review A}, 65(5):052112, 2002.

\bibitem{Munoz}
{Victoria Sanchez Munoz - TQC Conference}.
\newblock Chsh game with 3 players in a triangle with bi- and tri-partite
  entanglement.
\newblock
  \url{https://tqc-conference.org/wp-content/uploads/cfdb7_uploads/1687445204-poster-5787.pdf}.
\newblock School of Mathematical \& Statistical Sciences, Universtity of
  Galway.

\end{thebibliography}

\end{document}